\documentclass[correspondence]{IEEEtaes}

\usepackage{color,array}
\usepackage{graphicx}
\usepackage{cite}
\usepackage{float}
\usepackage{amsmath,amssymb,amsfonts}
\usepackage{textcomp}
\usepackage{mathbbol}
\usepackage{dirtytalk}
\usepackage{mathtools}
\usepackage[export]{adjustbox}[2011/08/13]
\usepackage{algorithm}
\usepackage{algorithmic}
\usepackage{comment}
\usepackage{amsthm}
\jvol{XX}
\jnum{XX}
\jmonth{XXXXX}
\paper{1234567}
\pubyear{2020}
\doiinfo{TAES.2020.Doi Number}
\newtheorem{theorem}{Theorem}
\newtheorem{lemma}{Lemma}

\newtheoremstyle{mydef}
  {0pt} 
  {0pt} 
  {\itshape} 
  {} 
  {\bfseries} 
  {.} 
  { } 
  {}
\theoremstyle{mydef}
\newtheorem{definition}{Definition}
\newtheorem{assumption}{Assumption}
\newtheorem{remark}{Remark}

\usepackage[font=footnotesize,labelfont=bf]{caption}
\usepackage{subcaption}
\usepackage{hyperref}
\usepackage{titlesec}

\titlespacing*{\section}
{0pt}      
{6pt}      
{3pt}      

\titlespacing*{\subsection}
{0pt}      
{4pt}      
{2pt}      

\titlespacing*{\subsubsection}
{0pt}
{3pt}
{1pt}

\titlespacing*{\section}{0pt}{6pt}{3pt}
\titlespacing*{\subsection}{0pt}{4pt}{2pt}
\begin{document}

\sptitle{Correspondence}
\title{Distributed Circumnavigation Using Bearing-Based Control with Limited Target Information}

\author{Kushal P. Singh}
\member{Student Member, IEEE}

\author{Manvi Bengani}

\author{Darshit Mittal}
\member{Member, IEEE}

 \author{Twinkle Tripathy}



\authoraddress{This work was not supported by any organisation. Kushal P. Singh (e-mail: kushalp20@iitk.ac.in) is a research scholar, M. Bengani (e-mail: manvib22@iitk.ac.in) is an undergraduate student, T. Tripathy (e-mail: ttripathy@iitk.ac.in) is an Assistant Professor with the Department of Electrical Engineering and D. Mittal (e-mail: darshitm23@iitk.ac.in) is an undergraduate student with the Department of Mechanical Engineering, Indian Institute of Technology Kanpur-208016, India.}


\markboth{CORRESPONDENCE}{}
\maketitle
\begin{abstract}
In this paper, we address the problem of circumnavigation of a stationary target by a heterogeneous group comprising of $\textbf{n}$ autonomous agents, having unicycle kinematics. The agents are assumed to have constant linear speeds, we control only the angular speeds. Assuming limited sensing capabilities of the agents, only a subset of agents, termed as \textit{leaders}, know the target location. The rest, termed as \textit{followers}, do not. We propose a distributed guidance law which drives all the agents towards the desired objective; global asymptotic stability (GAS) is ensured by using Zubov's theorem. The efficacy of the approach is demonstrated through both numerical simulations and hardware experiments.
\end{abstract}
%
\vspace{-0.1cm}
\section{Introduction}
Cooperative circumnavigation is fundamental in applications like surveillance, 
reconnaissance, environmental monitoring and multi-robot firefighting. Unicycle agents are widely used to achieve such tasks. They have inherent non-holonomic kinematic constraints and are under-actuated; yet, their use is motivated as they are simplified representations of several real-world vehicles.

When working as stand-alone systems, classical single-agent based circumnavigation approaches typically exploit range information\cite{milutinovic2017coordinate} or bearing information\cite{hashemi2015unmanned}. While single-agent solutions are useful, multi-agent coordination-based approaches offer scalability, fault tolerance, and reduced sensory requirements. Cyclic pursuit-based multi-agent approaches have been widely investigated, from early Lyapunov-based cyclic formations under limited communication \cite{limited_communication} to rotationally invariant formations involving heterogeneous agent speeds \cite{marshall2004formations,sinha2007generalization}. Recently, the authors in \cite{TRIPATHY2024111315} addressed local stability issues for unicycles with constant, identical linear speeds.

Further studies have explored more diverse formations. For agents modelled as integrators, the authors in \cite{ramirez2010distributed} propose strategies to achieve circular and elliptical formations. Distributed strategies for desired circular path generation using dynamic unicycles is explored in \cite{el2012distributed}. For heterogeneous agents, recent works have also focused on classifying motion regimes\cite{seyboth2014collective}, controlling only angular speeds or utilising attractive-repellent behaviours by controlling both linear and angular velocities\cite{zheng2015distributed}.
In scenarios with limited target information availability, the authors in \cite{yu2018circular,yu2018distributed} developed laws for cyclic and directed spanning-tree communication graphs among agents modelled as unicycles, succeeding even when one (root) agent tracks the target. 

Several of the afore-mentioned works require distance measurements or even provide local stability guarantees even by controlling both linear and angular speeds. In this paper, we address the problem of circumnavigation without distance measurements while also guaranteeing global system stability. Moreover, we control only the angular speeds of the agents with unicycle kinematics, thereby, facilitating an easier realworld implementation \cite{TRIPATHY2024111315}. Our major contributions are summarised as follows:
\vspace{-0.1cm}
\begin{itemize}
    \item \textit{Bearing based distributed guidance law:} We present a bearing based distributed guidance law that guarantees circumnavigation of a stationary target by a group of $n$ heterogeneous autonomous agents, by controlling only angular speeds. The proposed guidance law is distributed in the sense that the target's location is known only to a few agents. The use of bearing information and a memoryless design further simplifies its implementation.
    \item \textit{Global asymptotic stability:} The proposed distributed guidance law design ensures circumnavigation by the entire group from any arbitrary initial conditions. GAS is established using \textit{Zubov's theorem}\cite{khalil2002nonlinear}. 
    \item \textit{Hardware implementation:} The applicability of the proposed distributed guidance law is demonstrated through realworld experiments performed using TurtleBot robots.
\end{itemize}
\vspace{-0.1cm}
\vspace{-0.1cm}
\section{Preliminaries}
\label{sec:prelim}
\textit{Graph theory:} A directed graph $\mathcal{G}=(\mathcal{V},\mathcal{E})$, where $\mathcal{V}=\{1,2,\dots,n\}$ represents the set of nodes (agents) and $\mathcal{E}\subseteq \mathcal{V}\times \mathcal{V}$ represents the set of edges (communication links). The edge $(i,j) \in \mathcal{E}$ denotes an outgoing edge from node $i$ to node $j$; then, node $j$ becomes the out-neighbour of node $i$ and node $i$ the in-neighbour of node $j$. A node with no outgoing edges is called a {\it{sink node}}. Within this framework, the existence of such a link implies that agent $i$ can sense the heading angle of agent $j$ and the LOS angle from agent $i$ to $j$. 
%

\textit{Nonlinear systems:}  A nonlinear system is represented as $\dot{x}=f(t,x,u)$, where $f:D \to \mathbb{R}^n$ and $D$ is the domain. We refer to $x$ as the state, $t$ as time and $u$ as the input. It is referred to as an autonomous system if $\dot{x}=f(x)$. The standard definitions of stability, asymptotic stability and GAS are adopted from \cite{khalil2002nonlinear}.
\vspace{-0.2cm}
\begin{lemma}{(Zubov's Theorem)}
    \label{thm:zubov}
    Consider the non-linear system $\dot{x}=f(x)$ with an equilibrium at the origin and let $G \subset \mathbb{R}^n$ be a domain containing the origin. Suppose there exist two functions $V:G\rightarrow \mathbb{R}$ and $h:\mathbb{R}^n \rightarrow \mathbb{R}$ with the following properties: 
    \vspace{-0.1cm}
    \begin{itemize}
        \item $V$ is continuously differentiable and positive definite in $G$ and satisfies
        \begin{equation}
            \label{eq:cond_V}
            0<V(x)<1, \quad \forall x\in G \setminus\{0\}.
        \end{equation}
        \item As $x$ approaches the boundary of $G$, or in case of unbounded $G$ as $||x||\rightarrow\infty$, $\lim V(x)=1$.
        \item $h$ is continuous and positive definite on $\mathbb{R}^n$.
        \item For $x\in G$, $V(x)$ satisfies the partial differential eqn.
        \begin{equation}
        \label{eq:zubov_condition}
            ({\partial V}/{\partial x})f(x)=-h(x)[1-V(x)].
        \end{equation}
    \end{itemize}
    \vspace{-0.1cm}
    Then, $x=0$ is asymptotically stable and $G$ is the region of attraction (ROA).
\end{lemma}
\vspace{-0.35cm}
\begin{figure}[ht]
    \begin{center}
    \includegraphics[scale=0.75]{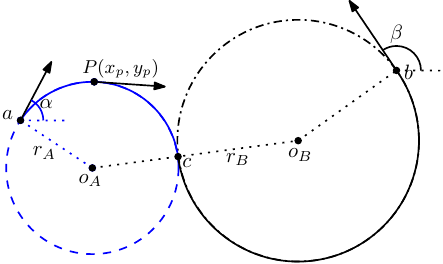}    
    \caption{Location of centres $\mathbf{o_A}$ and $\mathbf{o_B}$ of circles $\mathcal{C}_A$ and $\mathcal{C}_B$, respectively}  
    \label{fig:centers and radii}                                 
    \end{center}                                 
\end{figure}
\vspace{-0.45cm}

Consider the traversal problem of an autonomous vehicle starting and ending at pre-specified oriented points $A(\mathbf{a},\alpha)$ and $B(\mathbf{b},\beta)$, respectively. Suppose the traversal is along two circles $\mathcal{C}_A$ $\&$ $\mathcal{C}_B$ of radii $r_A$ $\&$ $r_B$, which are tangent to each other, as depicted in Fig. \ref{fig:centers and radii}. The coordinates of the centres $\mathbf{o_A}$ and $\mathbf{o_B}$ are given by $\mathbf{o_A}=(a_x+ r_{A}\sin{\alpha}, a_y - r_{A}\cos{\alpha})$ and $\mathbf{o_B}=(-r_{B}\sin{\beta}, r_{B}\cos{\beta})$, respectively. The specification of such a circle-circle (CC) trajectory is as given below:   
\vspace{-0.1cm}
\begin{lemma}[\cite{rao2024curvature}]
For any values of the oriented points $A(\mathbf{a},\alpha)$ and $B(\mathbf{b},\beta)$, a unique CC trajectory exists $\forall r_A\in\mathbb{R}$ such that $r_B$ is given by $r_B=p_3/(r_A-p_1)+p_2$ where:
\begin{align*}  
    p_1&=(a_x\sin\beta-a_y\cos\beta)/(1-\cos(\alpha-\beta)),\\
    p_2&=(a_x\sin\alpha-a_y\cos \alpha)/(1-\cos(\alpha-\beta)),\\
    p_3&=\left(\dfrac{a_x\sin{((\alpha+\beta)/2)}-a_y\cos{((\alpha+\beta)/2})}{1-\cos(\alpha-\beta)}\right)^2.
\end{align*}
\label{lem:feas_traj}
\end{lemma}
\vspace{-0.25cm}
\section{Problem Formulation}
\label{Section:Problem_Formulation}
\vspace{-0.25cm}
\begin{figure}[ht]
    \centering
        \begin{subfigure}[b]{0.23\textwidth}
        \centering    
        \includegraphics[width=0.8\linewidth,center]{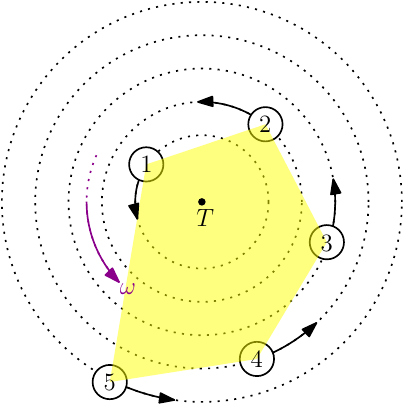}
        \caption{Desired circumnavigation.}
        \label{fig:desired_formation}
    \end{subfigure}
    \begin{subfigure}[b]{0.23\textwidth}
        \centering
        \includegraphics[width=\linewidth,center]{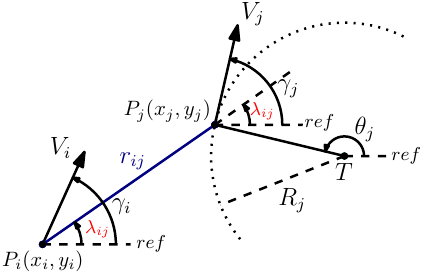}  
        \caption{Engagement geometry.}
        \label{fig:engage_geom}
    \end{subfigure}
    \caption{}
\end{figure}
\vspace{-0.1cm}
%
Circumnavigation refers to agents moving along circular paths around a central target, as shown in Fig.~\ref{fig:desired_formation}. Such motions are important in applications including surveillance, reconnaissance, environmental monitoring, and multi-robot firefighting. Motivated by these applications, we develop distributed guidance laws that enable autonomous vehicles to circumnavigate a fixed target.

We begin by considering a group of $n$ heterogeneous autonomous agents connected over a digraph $\mathcal{G(V,E)}$ and modelled as unicycles. A unicycle is a simplified representation of several real-world systems like cars, fixed-wing aircrafts, submarines, \textit{etc.} and, hence, are widely used in the guidance and control literature. The kinematics of such an agent is described by: 
\begin{equation}
\label{eq:unicycle_kinematics}
\dot{x_j} = v_j \cos \gamma_j, \quad
\dot{y_j} = v_j \sin \gamma_j, \quad
\dot{\gamma}_j = u_j,
\end{equation}

where $j\in\{1,2,...,n\}= \mathcal{V}$, $P_j(x_j,y_j)\in\mathbb{R}^2$ are the position coordinates, $\gamma_j\in \mathbb{S}^1$ is the its heading angle, $v_j\in\mathbb{R}$ is the constant forward speed and $u_j\in\mathbb{R}$ is the angular control input. In the given framework, we aim to achieve the desired objective by controlling only $u_j$ and keeping $v_j$ constant. This facilitates the practical implementability of the results. However, it makes the problem more challenging by further under-actuating the system. 
The spatial geometry in polar coordinates between any two agents $i$ and $j$ is given by:
    \begin{subequations}
    \label{eq:kinematics_polar}
    \begin{align}
    \label{eq:(r_ij_dot)}
    \dot{r}_{ij} & = v_j\cos{(\gamma_j-\lambda_{ij})}-v_i\cos{(\gamma_i-\lambda_{ij})}  \\
    \label{eq:(lambda_ji_dot)}
    r_{ij}\dot{\lambda}_{ij} &= v_j\sin{(\gamma_j-\lambda_{ij})}-v_i\sin{(\gamma_i-\lambda_{ij})},
    \end{align}
    \end{subequations}
where $r_{ij}$ is the distance between $i$ and $j$ and $\lambda_{ij}$ is LOS angle as shown in Fig. \ref{fig:engage_geom}. 

Without loss of generality  (WLOG), the target is assumed to be at the origin. In real-world scenarios, the sensing abilities of agents are often limited. Depending on individual sensing capabilities and initial positions, it is natural that not all the agents in the group know the target location initially. Keeping this in mind, we partition the agents into two disjoint sets: leaders ($\mathcal{L}$), who have target information, and the rest are followers ($\mathcal{F}$). 

With this, the objective can be formally stated as: given a group of agents connected over a digraph $\mathcal{G(V,E)}$ and limited target information availability, our aim is to design a distributed guidance law $u_j$ such that for any initial condition $(x_j(0), y_j(0), \gamma_j(0))$, each agent asymptotically circumnavigates the target. 
Mathematically, this requires the time-varying distance $r_j(t)$ to converge to the constant target radius $R_j$ for all $j \in \mathcal{V}$, expressed as $\lim_{t \to \infty} |r_j(t) - R_j| = 0$.
\section{Main Results}
\label{sec:Main_result}
    \begin{figure}[ht]
    \centering
    \includegraphics[scale=0.45]{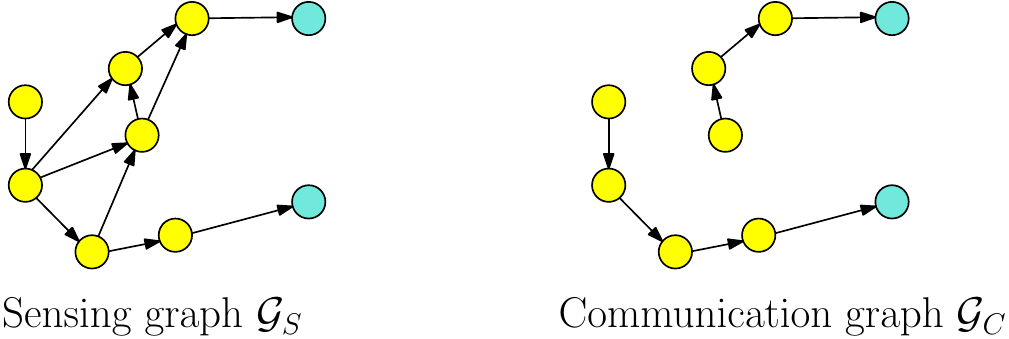}
    \caption{Leaders and followers are shown in blue and yellow, respectively.}
    \label{fig:sensing}
    \end{figure}
In this section, we present guidance laws to circumnavigate the target. To formalise the interaction structure among the agents, we introduce the following graph-theoretic definitions.
\begin{definition}[Sensing graph]
    \label{def:sensing_graph}
    We define the sensing graph $\mathcal{G}_S = (\mathcal{V}, \mathcal{E}_S)$. For nodes $i,j \in \mathcal{V}$, the existence of a directed edge $(i,j)\in \mathcal{E}_S$ implies that agent $i$ can sense agent $j$. 
\end{definition}
\begin{definition}[Communication graph]
    \label{def:communication_graph}
    The communication graph $\mathcal{G}_C=(\mathcal{V},\mathcal{E}_C)$ is a subgraph of $\mathcal{G}_S$. In this graph, each agent $i\in \mathcal{V}$ selects as its out-neighbour the closest agent it can sense in $\mathcal{G}_S$. If multiple nearest candidates exist, the agent may arbitrarily select any one of them (see Fig. \ref{fig:sensing}). Consequently, each follower node has exactly one out-neighbour. Additionally, the leaders are sink nodes. 
\end{definition}

Hence, $\mathcal{G}_C$ is not unique by construction and may contain multiple disconnected components even for a strongly connected $\mathcal{G}_S$. Further, we assume the following.

\begin{assumption}
    \label{assumption_mutli-out-neighb}
    In the communication graph $\mathcal{G}_C$, we assume that every follower $f \in \mathcal{F}$ has a directed path to at least one leader in the set $\mathcal{L}$.
\end{assumption}

As discussed earlier, the agents in the leader set $\mathcal{L}$ know the target location and can guide the followers towards it. Assumption \ref{assumption_mutli-out-neighb} ensures that the information from the leaders propagates to every follower in $\mathcal{F}$. With this, we now proceed for the guidance law design.
\subsection{Distributed guidance law for followers}
\label{subsec:follower_guidance_law}
In this subsection, we propose a distributed guidance law for all the followers in $\mathcal{F}$. WLOG, we assume that all the leaders follow suitable guidance laws and converge to stable circular trajectories around the target $T$ (see Fig. \ref{fig:engage_geom}). A suitable guidance law for leaders is discussed in the next subsection. 

Under Assumption \ref{assumption_mutli-out-neighb}, each agent possesses only one out-neighbour. Therefore, for any agent to successfully enter a circular path around the same centre as its out-neighbour (which can either be a leader or another follower) without using range information, the following set of geometrical conditions must be met, as stated next.
\begin{lemma}
\label{lemm:unique_equilibrium}
Consider any two agents $i,j\in \mathcal{V}$ such that $(i,j)\in\mathcal{E}_C$. The agents have unicycle kinematics given in eqn. \eqref{eq:unicycle_kinematics}. Suppose the agent $j$ traverses a fixed circular trajectory of radius $R_j$ (centred at the target $T$ with angular velocity $\omega_j$). If the following conditions are satisfied, then the agent $i$ is guaranteed to move on a concentric circle with an angular speed $\omega_j$.
\begin{enumerate}
    \item[a)] the heading angle difference remains constant, i.e. $ \dot{\gamma}_j(t) - \dot{\gamma}_i(t) = 0 $,
    \item[b)] the offset angle between the velocity vector and LOS remains constant, i.e. $\dot{\lambda}_{ij}(t)-\dot{\gamma}_i(t) = 0 $.
\end{enumerate}
\end{lemma}
%
\vspace{-0.3cm}
\begin{proof}
The position of agent $j$ can be expressed as $\overrightarrow{P_j}(t)=x_j(t)+ \iota y_j(t)$. Further, since agent $j$ is circumnavigating the target, its position vector can be written as $\overrightarrow{P_j}(t)=R_j \exp{(\iota\theta_j(t))}$ where the radius $R_j$ is constant and $\theta_j$ is the angular coordinate. With reference to Fig. \ref{fig:engage_geom}, it follows from the circular geometry that $\gamma_j(t)=\theta_j(t)-\pi/2$. Its velocity vector is then given by $\overrightarrow{v_j}(t)=\frac{d}{dt}\overrightarrow{P_j}(t)=\iota R_j\dot{\theta}_j(t)\exp{(\iota\theta_j(t))}$. A constant $v_j$ necessitates that the $\dot{\theta}_j(t)$ is also constant.

Now we analyse the relative motion between the two agents under conditions (a) and (b) of the lemma statement, and we suitably define two error variables for  $i$:
    \begin{subequations}
    \label{eq:error variables}
    \begin{align}
    \label{eq:e_1^j}
     e_{i_1} &\triangleq \gamma_j - \gamma_i \\
    \label{eq:e_2^j}
     e_{i_2} &\triangleq \lambda_{ij} - \gamma_i
    \end{align}
    \end{subequations}

First, we analyse the implication of condition (a). We start by substituting the condition $\gamma_i(t) = \gamma_j(t) - e_{i_1}$ into the expression for $\overrightarrow{v_i}(t)$ and simplifying, we get $\overrightarrow{v_i}(t)=(v_i/v_j)(\exp{(-\iota e_{i_1}}))\overrightarrow{v_j}(t)$.
So, $\overrightarrow{v_i}(t)$ is proportional to $\overrightarrow{v_j}(t)$ by a constant, $\alpha \triangleq (v_i/v_j)\exp{(-\iota e_{i_1})}$. Hence, exploiting the velocity and position relationship, we get $\overrightarrow{P_i}(t)=\alpha \overrightarrow{P_j}(t)+c$, where $c$ is a constant of integration. Further, $\overrightarrow{P_i}(t)=\alpha R_j\exp{\iota\theta_j(t)}+c$ as $\overrightarrow{P_j}(t)=R_j \exp{(\iota\theta_j(t))}$. Note that the equation describes a circular path for agent $i$ of radius $|\alpha|R_j$ and centred at the point $c$.
Next, on applying condition (b), the angular separation $e_{i_2}$ (as defined in eqn. \eqref{eq:e_2^j}) is constant. The LOS vector from $i$ to $j$ is $\overrightarrow{r_{ij}}(t) = \overrightarrow{P_j}(t)-\overrightarrow{P_i}(t)$, which is given by:
\begin{equation}
\label{eq:integrated_P_j}
\overrightarrow{r_{ij}}(t) = \overrightarrow{P_j}(t) - (\alpha\overrightarrow{P_j}(t)+c) = (1-\alpha)\overrightarrow{P_j}(t)-c.
\end{equation}
Expressed as a function of $\theta_j$, eqn. \eqref{eq:integrated_P_j} is $\overrightarrow{r_{ij}}(\theta_j) = (1-\alpha)R_j\exp{(\iota\theta_j(t))}-c$.
Then, condition (b), combined with $\gamma_j(t)=\theta_j(t)-\pi/2$, implies that $\arg{(\overrightarrow{r_{ij}}(\theta_j))} - \theta_j$ must be a constant. For this relationship to hold, its derivative w.r.t. $\theta_j$ must be zero:
\begin{equation}
\label{eq:lead_angle_condition}
    \dfrac{d}{d\theta_j}(\arg{\overrightarrow{r_{ij}}(\theta_j(t))} - \theta_j(t)) = 0.
\end{equation}
%
Using the identity for the derivative of an argument, $\frac{d}{d\theta_j}(\arg{f(\theta_j)})=\text{Img}\left(f'(\theta_j)/f(\theta_j)\right)$ , we get:
\begin{equation}
\label{eq:zero_condition}
\text{Img}\{(\overrightarrow{r_{ij}}'(\theta_j))/(\overrightarrow{r_{ij}}(\theta_j))\} - 1 = 0.
\end{equation}
We compute the derivative $\overrightarrow{r_{ij}}'(\theta_j)$ and the ratio becomes:
\begin{equation}
    \label{eq:dash_by_normal}
    \dfrac{\overrightarrow{R_{ij}}'(\theta_j)}{\overrightarrow{R_{ij}}(\theta_j)}=\dfrac{\iota(1-\alpha)R_j\exp{(\iota\theta_j})}{(1-\alpha)R_j\exp{(\iota\theta_j)}-c}.
\end{equation}
Substituting the result of the eqn. \eqref{eq:dash_by_normal} in the condition from the eqn. \eqref{eq:zero_condition}, we get:
\begin{equation}
\label{eq:arg_img}
 \text{Img}\left(\dfrac{\iota(1-\alpha)R_j\exp{(\iota\theta_j)}}{(1-\alpha)R_j\exp{(\iota\theta_j)}-c}\right) -1 = 0.  
\end{equation}
 If we test $c=0$, the expression simplifies to $\text{Img}(\iota) - 1 = 1-1 = 0$, which is true. If $c \neq 0$, the term in the eqn. \eqref{eq:arg_img} does not simplify to zero. Therefore, the only possible solution that satisfies the condition for all time is $c=0$.

Since $c=0$, the trajectory for $i$ is $\overrightarrow{P_i}(t)=\alpha\overrightarrow{P_j}(t)$. This confirms that agent $i$ circumnavigates the same centre $O$ as agent $j$ and shares the same angular velocity $\omega_j$.
\end{proof}
\begin{remark}
    \label{rem:rigid_body_rem}
    When Lemma \ref{lemm:unique_equilibrium} holds, the triangle $\triangle TP_iP_j$ (see Fig. \ref{fig:engage_geom}) is a rigid body. Its side lengths, the radii $R_j = \|P_j(t) - O\|$ and $R_i = \|P_i(t) - O\|$, and the inter-agent distance $r_{ij}$ are all invariant for all $ t \in \mathbb{R} $.
\end{remark}

With the required information to circumnavigate, we move on to its implementation in the following result.
\begin{theorem}
    \label{thm:conv_sing_fol}
    Consider a system of $n$ agents governed by kinematics defined in eqn. \eqref{eq:unicycle_kinematics}. Let every leader $l \in \mathcal{L}$ move on a stable circular path such that $\lim_{t\to t_f} r_l(t) = \bar{r}_{l}$ and $\dot\gamma_l(t)\to \bar{\omega}_{l}$. Under Assumption \ref{assumption_mutli-out-neighb} and $C_1,C_2 \in \mathbb{R}^+$, each follower $i \in \mathcal{F}$ is guaranteed to circumnavigate the target $T$ at the same angular rate as the leader at the terminus of its directed path in $\mathcal{G}_C$ under the following distributed guidance law:
    \begin{equation}
    \label{eq:follower_guidance}
    \dot{\gamma}_i = C_1(\gamma_j - \gamma_i) + C_2 \sin(\lambda_{ij} - \gamma_i).
    \end{equation}
\end{theorem}
\begin{proof}
The proof proceeds in two stages: first, establishing the stability of a single follower with an out-neighbour that circumnavigates the target, and second, propagating this stability through the communication graph via induction.

 Consider a single follower $i \in \mathcal{F}$ having an out-neighbour $j \in \mathcal{L}$ that is already on a stable circular path (see Fig. \ref{fig:engage_geom}). Now based on the equilibrium conditions from Lemma \ref{lemm:unique_equilibrium}, the two error variables $e_{i_1} = \gamma_j - \gamma_i$ and $e_{i_2} = \lambda_{ij} - \gamma_i$ are constant at equilibrium. Let the state vector be $\mathbf{e}_i = [e_{i_1} \quad e_{i_2}]^T$. The system dynamics are given by:
\begin{equation}
\label{eq:m_dot}
\dot{\mathbf{e}}_i = f(\mathbf{e}_i) = [\dot{e}_{i_1} \quad \dot{e}_{i_2}]^T = [f_1 \quad f_2]^T.
\end{equation}

Substituting the values of $\dot{e}_{i_1}$ and $\dot{e}_{i_2}$ in eqn. \eqref{eq:m_dot} gives: 
\begin{equation}
\label{eq:f(e)}
f(\mathbf{e}_i) = \begin{bmatrix} 
\omega_j - \dot{\gamma}_i \\
\dfrac{v_j \sin(e_{i_1} - e_{i_2}) + v_i \sin(e_{i_2})}{r_{ij}} - \dot{\gamma}_i
\end{bmatrix},
\end{equation}
where $\dot{\gamma}_i=C_1 e_{i_1} + C_2 \sin(e_{i_2})$
Let the equilibrium value of $\mathbf{e}_i$ be $\bar{\mathbf{e}}_{i}$. The equilibrium values $\bar{e}_{i_1}$ and $\bar{e}_{i_2}$ in eqn. \eqref{eq:f(e)} gives the condition:
\begin{equation}
    \label{eq:eq_cond}
    \omega_j = C_1 \bar{e}_{i_1} + C_2 \sin(\bar{e}_{i_2}).
\end{equation}

With the equilibrium defined, we proceed to analyse its stability properties. We employ Zubov's Theorem for this purpose, as it allows the simultaneous characterisation of the ROA and the stability of the origin. Accordingly, we define a new state variable $\mathbf{z}_i=\mathbf{e}_i-\bar{\mathbf{e}}_{i}$ to shift equilibrium to origin, where $\mathbf{z}_i=[z_{i_1} \quad z_{i_2}]^T$. The new system dynamics are $\dot{\mathbf{z}}_i\triangleq\tilde{f}(\mathbf{z}_i)=\dot{\mathbf{e}}_i=f(\mathbf{z}_i+\bar{\mathbf{e}}_{i}).$

Substituting $e_{i_1}=z_{i_1}+\bar{e}_{i_1}$, $e_{i_2}=z_{i_2}+\bar{e}_{i_2}$, and $\omega_j$ (from eqn. \eqref{eq:eq_cond}) into eqn. \eqref{eq:f(e)}, we obtain $\tilde{f}(\mathbf{z}_i)
=
\begin{bmatrix}
\tilde{f}_{i_1} \quad \tilde{f}_{i_2}
\end{bmatrix}^T,$ where $\tilde{f}_{i_1}= -C_1 z_{i_1}
 + C_2( \sin(\bar{e}_{i_2})
 - \sin(z_{i_2}+\bar{e}_{i_2}))$ and $\tilde{f}_{i_2}=
 (v_j \sin(z_{i_1}-z_{i_2}+\bar{e}_{i_1}-\bar{e}_{i_2})
 + v_i \sin(z_{i_2}+\bar{e}_{i_2})
 )/r_{ij}- C_1 (z_{i_1}+\bar{e}_{i_1})
 - C_2 \sin(z_{i_2}+\bar{e}_{i_2})$.
We choose functions $V(\mathbf{z}_i)=1-\exp{(-\alpha((z_{i_1})^2+(z_{i_2})^2))}$, where $\alpha \in \mathbb{R}^+$, and $h(\mathbf{z}_i)=-2\alpha[z_{i_1} \quad z_{i_2}]\tilde{f}(\mathbf{z}_i)$. These functions satisfy the conditions laid down in Lemma \ref{thm:zubov} in Section \ref{sec:prelim}. The function $V$ is continuously differentiable, positive definite in $\mathbb{R}^2$ ($G=\mathbb{R}^2$), and satisfies eqn. \eqref{eq:cond_V}. Furthermore, the limit of $V$ as $||\mathbf{z}^i||\rightarrow \infty$ is $1$.
The constants $C_1$ and $C_2$ are selected to ensure that $h$ is positive definite. We then verify the condition given in eqn.~\ref{eq:zubov_condition} w.r.t $\mathbf{z}_i$.

For this purpose, we substitute the partial derivative of $V(\mathbf{z}_i)$ and the system dynamics $\tilde{f}(\mathbf{z}_i)$ into the left-hand side of eqn. \eqref{eq:zubov_condition}. We obtain:
\begin{align} 
 \frac{\partial V}{\partial \mathbf{z}_i}\tilde{f}(\mathbf{z}_i)&= \underbrace{2\alpha e^{-\alpha ((z_{1i})^2)+(z_{2i})^2)} [z_{i_1} \quad z_{i_2}]}_{\partial{V}/\partial{\mathbf{z}_i}}\underbrace{[\tilde{f}_{i_1} \quad \tilde{f}_{i_2}]^T }_{\tilde{f}(\mathbf{z}_i)}  \notag \\  &= 2\alpha e^{-\alpha ((z_{i_1})^2)+(z_{i_2})^2)} \left( z_{i_1} \tilde{f}_{i_1} + z_{i_2} \tilde{f}_{i_2} \right). \label{eq:derivation_step2} 
\end{align}

Recalling proposed $h(\mathbf{z}_i)$ from above and observing that $1-V(\mathbf{z}_i) = e^{-\alpha ((z_{i_1})^2)+(z_{i_2})^2)^2}$, we can rewrite eqn. \eqref{eq:derivation_step2} as $-h(\mathbf{z}_i) [1-V(\mathbf{z}_i)]$, which verifies eqn. \eqref{eq:zubov_condition}.

As per Lemma \ref{thm:zubov} stated in Section \ref{sec:prelim}, the origin is asymptotically stable. Additionally, the region of attraction is $\mathbb{R}^2$. Hence, the origin is GAS. Next, we prove the stability of the entire group.

Now, to prove the stability of the entire group. We define the radial error $d_{j_r}(t) \triangleq r_j(t)-{R}_j$, where $r_j(t)$ is the distance from the target and $R_j$ is a constant value. Leaders $l \in \mathcal{L}$ converge by definition in theorem statement, so $\lim_{t\to t_f} d_{l_r}(t) = 0$ and $\dot\gamma_l \to \bar{\omega}_l$.
Now, consider a path $l_m \leftarrow f_1 \leftarrow \dots \leftarrow f_k$. Assume follower $f_k$ has converged to a stable circular path with speed $\bar{\omega}_{l_m}$. For the next follower $f_{k+1}$, the agent $f_k$ acts as a stable leader. Applying the result from stage one of the proof, $f_{k+1}$ must also converge to a circular path and match the angular speed of $f_k$ (and thus of $l_m$). 
By Assumption \ref{assumption_mutli-out-neighb}, every follower has a path to a leader, this inductive argument cascades to all $f \in \mathcal{F}$. Hence, all agents converge to their respective radii and angular speeds.
\end{proof}

This concludes the analysis for followers. Since the leaders' circumnavigation task is simpler due to more information, their approach is discussed after the followers' in the next subsection.
\subsection{Guidance law for leaders}
\label{subsec:guidance_law_leaders}
We recall that the leaders have information about the target and aim to circumnavigate it, as do the followers. In the proposed design methodology for the followers, the leaders in $\mathcal{L}$ are assumed to follow circular trajectories. Keeping practical implementability in mind, we employ a guidance law for leaders that drives them toward suitable trajectories within a feasible, prescribed time.

We employ CC trajectories as they offer sufficient control over $\beta$ and $r_B$ (see Fig. \ref{fig:centers and radii}) and enable leaders to autonomously select $\alpha$, $ \beta$, and $r_B$, thereby facilitating collision avoidance. For the same, we leverage a geometric result with constant inputs from our prior work in \cite{rao2024curvature} to implement these trajectories (see Lemma \ref{lem:feas_traj}). This formulation ensures that for any feasible $r_A$, a unique $r_B$ exists to connect $\mathbf{a}$ and $\mathbf{b}$ via two tangent circular arcs. Consequently, any initial configuration can be steered to the desired target path using only two circular arcs.

Having concluded the theoretical analysis and proofs, we now proceed to validate these findings. 
\section{Simulation Results}
\label{Sec:Simulation}
In this section, we validate the proposed guidance law using two static graph topology setups: a single-leader case and a two-leader case. In all plots, leaders are represented by blue and followers by yellow.
%
\begin{table}[h]
\caption{Simulation's initial conditions (in SI units).}
\centering
\begin{tabular}{|c|c|c|c|c|c|c|}
\hline
 & \multicolumn{3}{c|}{\text{Case 1}} & \multicolumn{3}{c|}{\text{Case 2}} \\
\hline
Agent 
& $P$ & $V$ & $\gamma$ & $P$ & $V$ & $\gamma$ \\
\hline
1 & (2,2)     & 18 & -2.44 & (2,1)    & 28 & -2.53\\
\hline
2 & (-24,-15) & 36 & -1.48  & (34,-5)  & 35 & -1.57\\
\hline
3 & (-20,35)  & 74 & 0.61   & (-10,15) & 50 & -0.52\\
\hline
4 & (30,-45)  & 99 & -1.19  & (10,-25) & 18 & -1.05\\
\hline
5 & (-29,-11) & 24 & -1.34  & (-25,-15)& 22 & -1.31\\
\hline
6 & NA         & NA  & NA             & (10,23)  & 40 & -1.43\\ 
\hline
\end{tabular}
\label{tab1}
\end{table}
\subsection*{Case $1$: Single leader configuration}
\begin{figure}[ht]
    \centering
    \begin{subfigure}[b]{0.23\textwidth}
        \centering
        \includegraphics[width=0.7\linewidth,center]{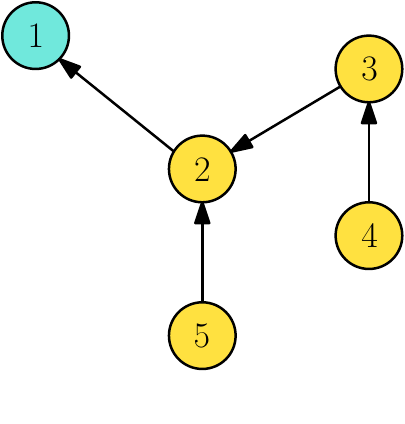}  
        \caption{Communication graph.}
        \label{fig:first_graph}
    \end{subfigure}
    \begin{subfigure}[b]{0.23\textwidth}
        \centering    
        \includegraphics[width=\linewidth,center]{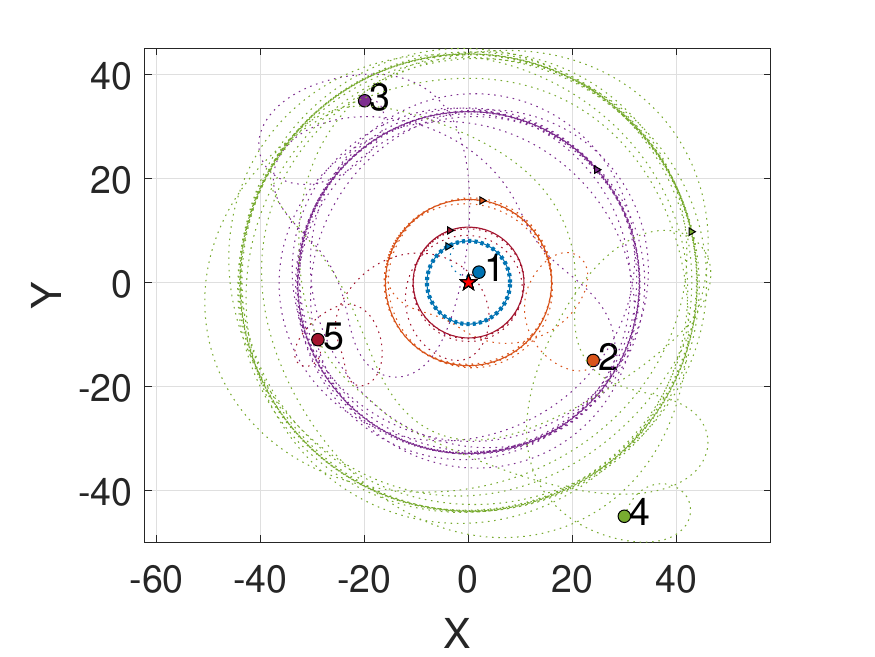}
        \caption{Trajectories}
        \label{fig:trajectory_case1}
    \end{subfigure}
    \begin{subfigure}[b]{0.23\textwidth}
        \centering
        \includegraphics[width=\linewidth,center]{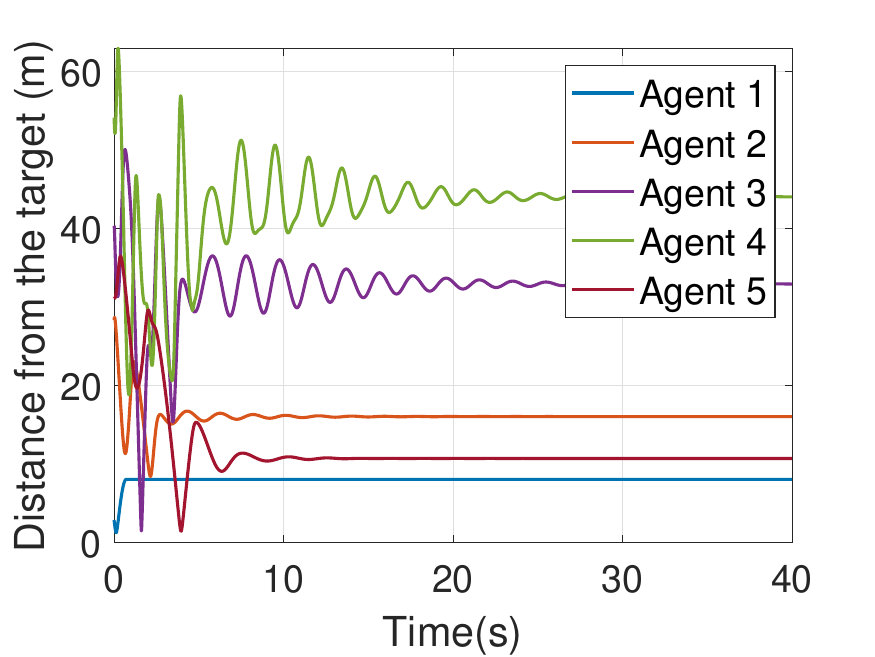}  
        \caption{Distances from the target}
        \label{fig:distance_from_target_case1}
    \end{subfigure}
    \begin{subfigure}[b]{0.23\textwidth}
        \centering    
        \includegraphics[width=\linewidth,center]{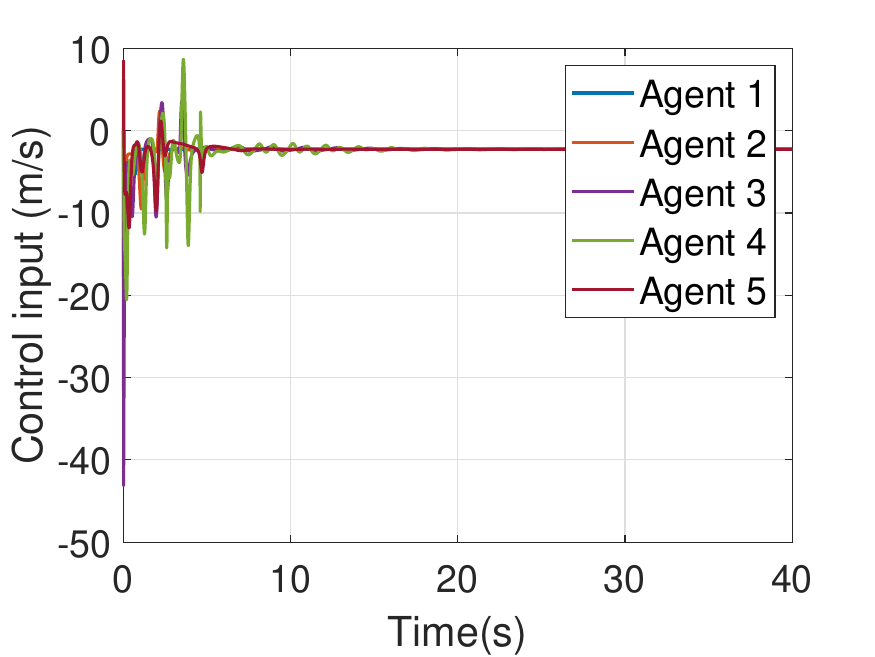}
        \caption{Control inputs.}
        \label{fig:control_inputs_case1}
    \end{subfigure}
    \caption{Case $1$.}
    \label{fig:sim_one}
\end{figure}

Fig. \ref{fig:first_graph} depicts the communication topology for five agents. Consistent with Theorem \ref{thm:conv_sing_fol}, agents converge to circular orbits (Fig. \ref{fig:trajectory_case1}) with stabilised horizontal distances (Fig. \ref{fig:distance_from_target_case1}). Moreover, Fig. \ref{fig:control_inputs_case1} shows all control inputs converging to a unified value, confirming the formation of a single inavriant rotating structure.
\subsection*{Case $2$: Two leader configuration}
\begin{figure}[ht]
    \centering
    \begin{subfigure}[b]{0.23\textwidth}
        \centering
        \includegraphics[width=0.7\linewidth,center]{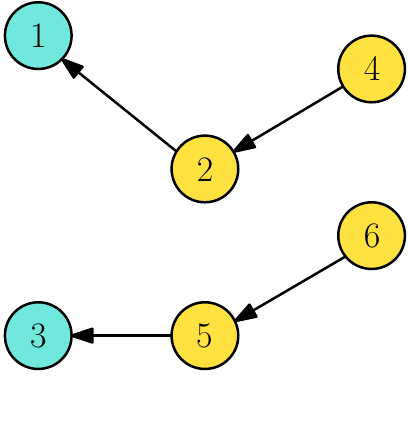}  
        \caption{Communication graph}
        \label{fig:second_graph}
    \end{subfigure}
    \begin{subfigure}[b]{0.23\textwidth}
        \centering    
        \includegraphics[width=\linewidth,center]{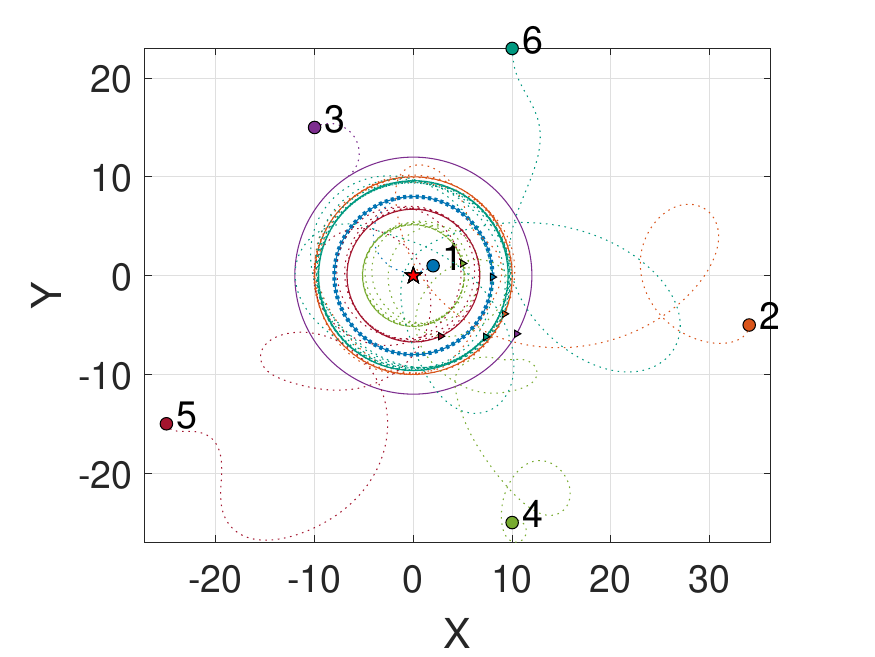}
        \caption{Trajectories}
        \label{fig:trajectory_case2}
    \end{subfigure}
    \begin{subfigure}[b]{0.23\textwidth}
        \centering
        \includegraphics[width=\linewidth,center]{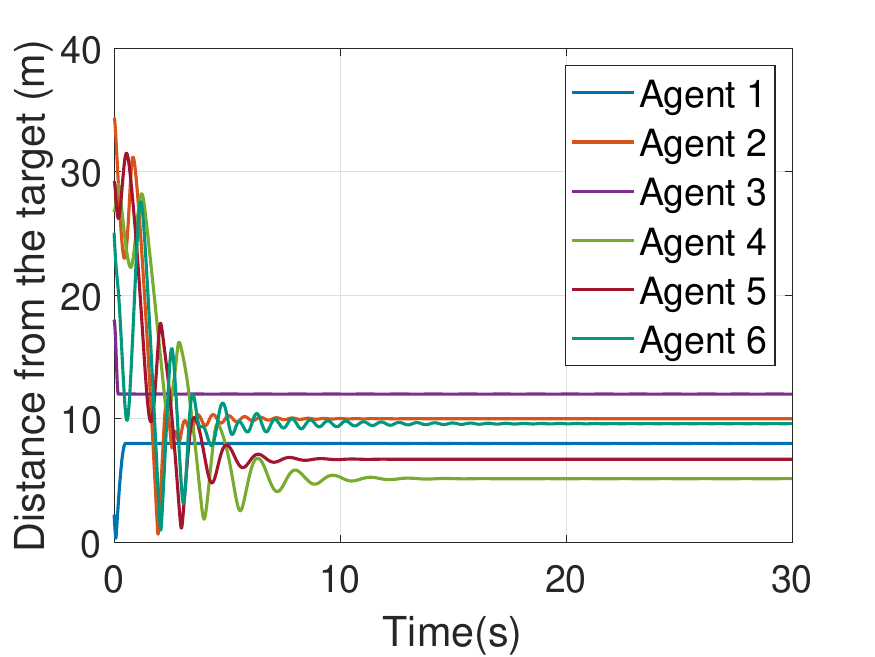}  
        \caption{Distances from the target}
        \label{fig:distance_from_target_case2}
    \end{subfigure}
    \begin{subfigure}[b]{0.23\textwidth}
        \centering    
        \includegraphics[width=\linewidth,center]{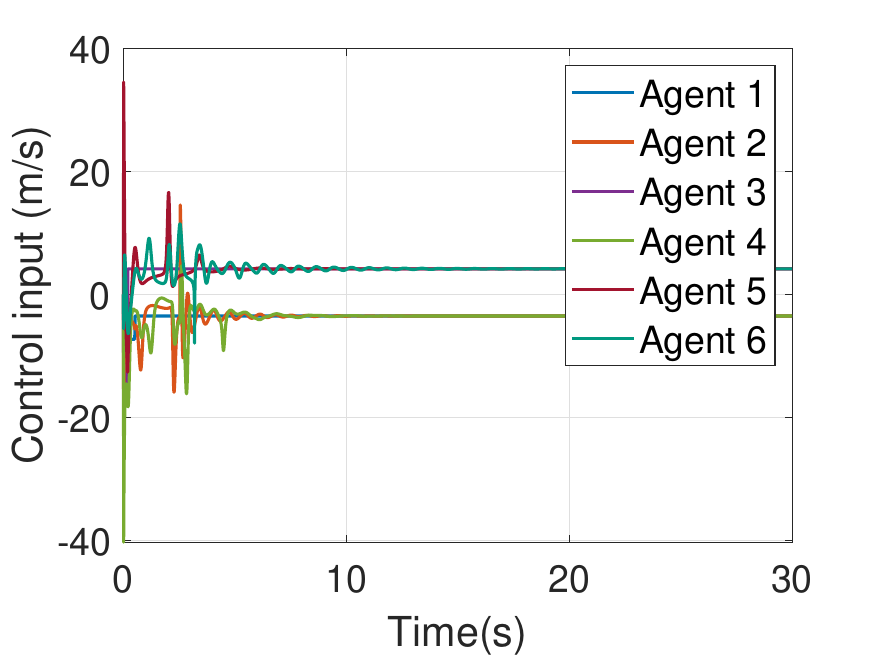}
        \caption{Control inputs}
        \label{fig:control_input_case2}
    \end{subfigure}
    \caption{Case $2$.}
    \label{fig:sim_two}
\end{figure}

Fig. \ref{fig:second_graph} shows the topology for a six-agent, two-leader system. As predicted, agents converge to circular paths (Fig. \ref{fig:trajectory_case2}) with constant target separation (Fig. \ref{fig:distance_from_target_case2}). Control inputs (Fig. \ref{fig:control_input_case2}) reveal that $\omega_2, \omega_3$ synchronize with $\omega_1$, while $\omega_4, \omega_5$ synchronize with $\omega_6$. This results in two distinct invariant rotating bodies: a triangle connecting agents $1\text{--}3$ and another connecting agents $4\text{--}6$. These simulations confirm theoretical predictions.
\section{Experimental Validation}
\label{sec:experiments}
In this section, we demonstrate practical feasibility through hardware experiments with a leader-follower pair of ground mobile robots. These tests validate theoretical predictions and assess the robustness against real-world challenges, including sensor noise, communication delays, and actuation constraints.
The testbed comprises two infrared reflective marker-equipped TurtleBot3 Burger robots ($v_{max}=0.22\text{m/s}$). They are tracked by a six-camera OptiTrack system providing sub-millimetre accuracy at $120\text{Hz}$. A centralised node computes relative angles ($\gamma_j - \gamma_i$ and $\lambda_{ji} - \gamma_i$) from the tracking data and transmits commands via WiFi. To ensure faster convergence, the leader was pre-positioned on a 0.7 m radius circle, while the follower started from an arbitrary condition.
The leader's and follower's linear speeds were set to 0.105\,m/s and 0.15\,m/s, respectively. Their respective initial positions were $(0.7\,\text{m}, 0\,\text{m})$ and $(0.87\,\text{m}, 0.16\,\text{m})$, with initial headings of $90^\circ$ and $139^\circ$.
\begin{figure}[ht]
    \centering
    \begin{subfigure}[b]{0.23\textwidth}
        \centering
        \includegraphics[width=\linewidth,center]{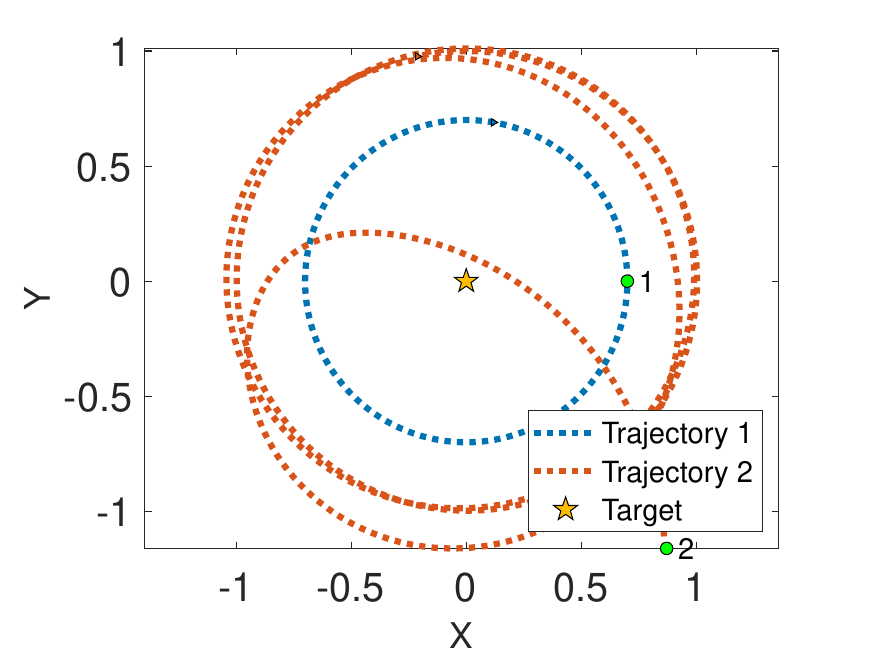}  
        \caption{Simulation trajectory}
        \label{fig:traj_exp_simu}
    \end{subfigure}
    \begin{subfigure}[b]{0.23\textwidth}
        \centering    
        \includegraphics[width=\linewidth,center]{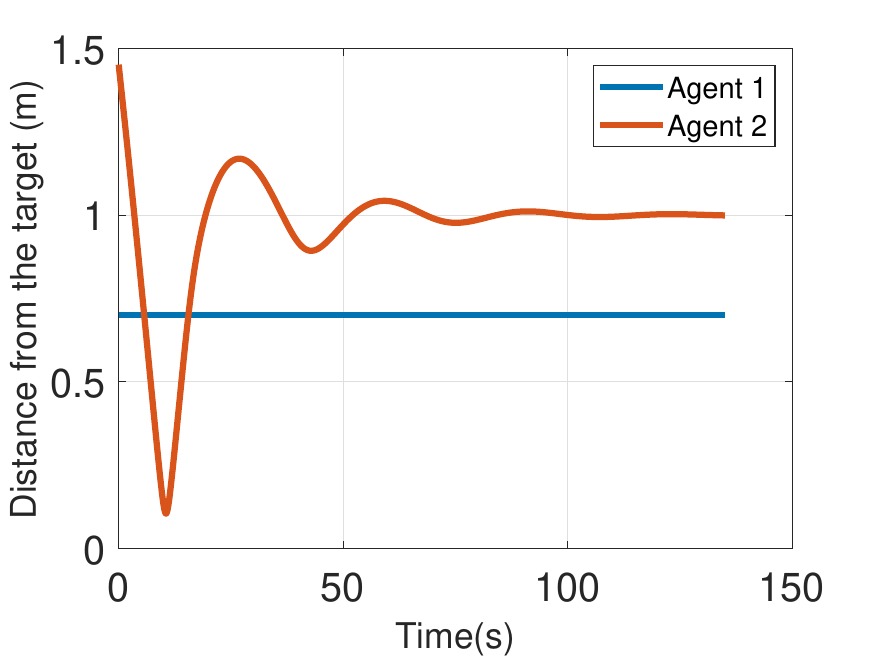}
        \caption{Simulation distance}
        \label{fig:dist_exp_simu}
    \end{subfigure}
    \begin{subfigure}[b]{0.23\textwidth}
        \centering
        \includegraphics[width=\linewidth,center]{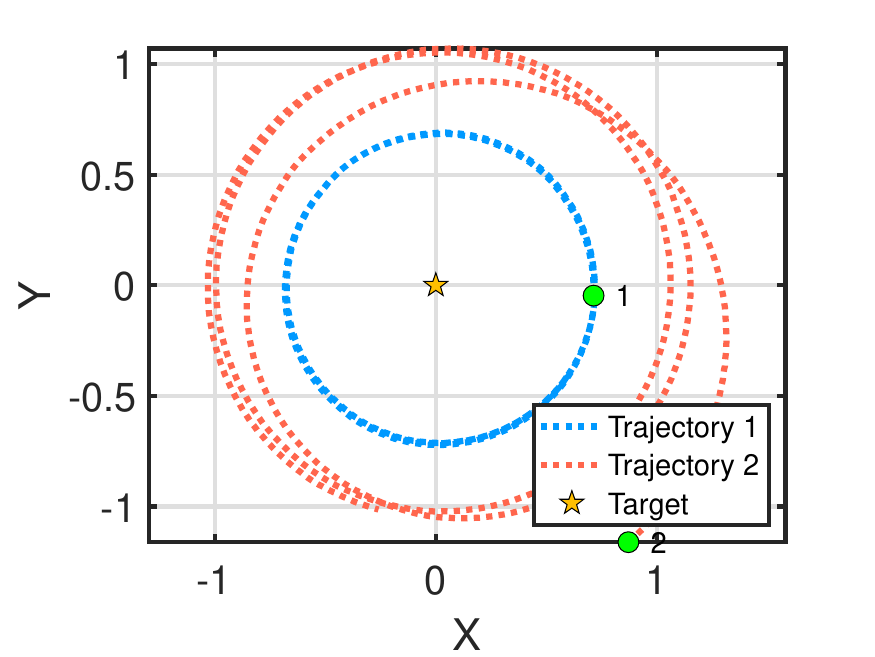}  
        \caption{Experimental trajectory}
        \label{fig:traj_exp}
    \end{subfigure}
    \begin{subfigure}[b]{0.23\textwidth}
        \centering    
        \includegraphics[width=\linewidth,center]{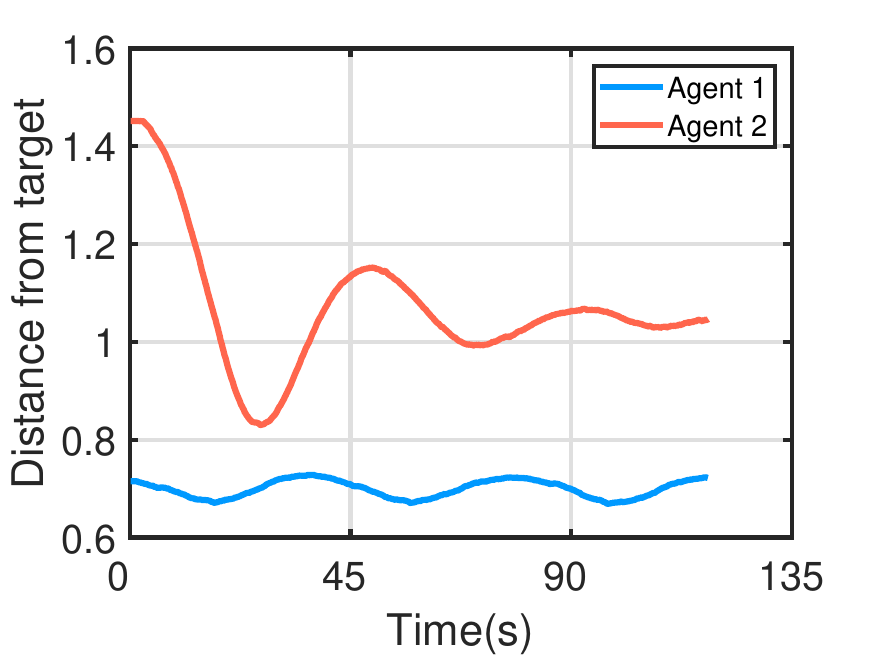}
        \caption{Experimental distance}
        \label{fig:dist_exp}
    \end{subfigure}
    \caption{Comparison of simulation and experimental results.}
    \label{fig:exp_one}
\end{figure}

In Fig. \ref{fig:exp_one} (Agent 1: leader, Agent 2: follower), trajectories in both simulation (Fig. \ref{fig:traj_exp_simu}) and experiment (Fig. \ref{fig:traj_exp}) confirm that the follower successfully achieves concentric circular motion. While the simulated follower settles at $0.999\,\text{m}$ (Fig. \ref{fig:dist_exp_simu}), experimental distance settles between $0.96\,\text{m}$ and $1.04\,\text{m}$ (Fig. \ref{fig:dist_exp}). The leader's experimental distance exhibits noise induced waviness ($0.66\text{--}0.74\,\text{m}$) around the $0.7\text{m}$ target, but remains bounded. Due to space constraints, only the faster-follower scenario is presented. The video link for the experiment is \href{https://youtu.be/-4YHJBETVX0}{https://youtu.be/-4YHJBETVX0}.
\section{CONCLUSION}
\label{Sec:Conclusion}
In this paper, we present a bearing based distributed guidance law to achieve circumnavigation of a stationary target by a group of $n$ heterogeneous autonomous agents, modelled as unicycles. The agents are assumed to have constant linear speeds; this further under-actuated the model but enhances its applicability. In the given framework, only a subset of the agents know the target location, referred to as \textit{leaders} while the rest are called \textit{followers}. For the leaders, we employ circular trajectories for ensuring finite-time target circumnavigation. A distributed bearing-based guidance law is designed for the followers, as presented in Theorem \ref{thm:conv_sing_fol}. The proposed design ensures GAS even by controlling only the angular speeds. The results are illustrated through numerical simulations. The practical implementability has also been demonstrated through realworld hardware experiments performed on TurtleBots using an OptiTrack motion capture system.

\correspauthor%

\bibliographystyle{IEEEtran}  
\bibliography{citation}

\end{document}